\documentclass[12pt]{iopart}
\usepackage{graphicx}
\begin{document}
\title[Quantum entanglement of spin-1 bosons]{Quantum entanglement
of spin-1 bosons with coupled ground states in optical lattices}

\author{B \"Oztop$^1$, M \"O Oktel$^1$,
\"{O} E M\"{u}stecapl\i o\u{g}lu$^2$ and L You$^3$}

\address{$^1$ Department of Physics, Bilkent University, Ankara
06800, Turkey}
\address{$^2$ Department of Physics, Ko\c{c} University, Sariyer 34450,
Istanbul, Turkey}
\address{$^3$ School of Physics, Georgia Institute of Technology, Atlanta,
Georgia 30332, USA}

\ead{boztop@fen.bilkent.edu.tr}

\begin{abstract}
We examine particle entanglement, characterized by pseudo-spin
squeezing, of spin-1 bosonic atoms with coupled ground states in a
one-dimensional optical lattice. Both the superfluid and
Mott-insulator phases are investigated separately for
ferromagnetic and antiferromagnetic interactions. Mode
entanglement is also discussed in the Mott insulating phase. The
role of a small but nonzero angle between the polarization vectors
of counter-propagating lasers forming the optical lattice on
quantum correlations is investigated as well.
\end{abstract}
\pacs{03.75.Lm, 03.75.Mn, 03.67.Bg} \submitto{\JPB} \maketitle

\section{Introduction}
\label{sec:intro}
The investigation of atomic bosons with short-range repulsive
interactions in a periodic potential by using the Bose-Hubbard
model has revealed a quantum phase transition between two distinct
phases: a superfluid and a Mott-insulator, that exists at
sufficiently low temperatures \cite{Fisher-PRB89}. The formalism
of the Bose-Hubbard model was successfully mapped onto a system of
cold bosonic atoms in an optical lattice \cite{Zoller-PRL98}. The
superfluid to Mott insulator phase transition was experimentally
realized \cite{Bloch-Nature02} and further examined and
theoretically digested \cite{Ho-PRL07}. Continued progresses have
focused on systems of multi-component Bose-Einstein condensates
(BECs) in an optical lattice \cite{Wu-PRA03}, where diverse topics
such as quantum phase transitions of spin-2 bosons
\cite{Ge-PRA03}, two-component condensates \cite{KG-PRL03}, and
spin-1 bosons with coupled ground states \cite{KG-PRA04} are
studied.

An interesting feature characterizing a variety of lattice models
mapped onto atomic gases is quantum entanglement. Additionally,
cold atom based lattice models have been identified as ideal
candidates for universal quantum emulation of strongly interacting
many body systems. While a complete understanding of quantum
entanglement and correlations in an atomic lattice model remains a
significant challenge even in theoretical terms
\cite{Amico-review}, much has been understood for an important
type of correlation, the so-called spin squeezing, or pseudo spin
squeezing. For those systems that undergo quantum phase
transitions, the presence and the measure of entanglement is
important not only at the transition point, but also for the
different phases of the system. These systems show various
behaviors, entanglement and disentanglement, coherent and squeezed
spin states, mode and particle entanglement for different phases
that can be controlled by interaction types and strengths as well
as lattice configurations.

Squeezed spin states are states whose spin fluctuation in one of
the transverse spin components is below the standard quantum
limit. It was shown in Ref.~\cite{Kitagawa-PRA93} a spin-$s$
squeezed spin state is a correlated state consisting of $2s$
spin-$1/2$ particles. This implies a potential connection between
spin squeezing and entanglement, due to the existence of
correlations affecting the separability of a system with many
spin-$1/2$ particles \cite{Zoller-Nature01}. Spin squeezing can
occur in many models with a variety of atom-atom interactions
\cite{Helmerson-PRL01,Yi-PRA06}, for atomic condensates inside
external traps \cite{Zoller-Nature01}, and for atoms inside
optical lattices \cite{Molmer-PRL99}.

In this work, we are interested in the possibility and the
condition for spin squeezing in the pseudo-spin of coupled ground
states in an optical lattice model with spin-1 bosons. We hope to
explore spin squeezing properties of the system carefully studied
in Ref.~\cite{KG-PRA04}. This paper is organized as follows. In
Sec.~\ref{s:model}, we review the model system
\cite{KG-PRA04,KG-PRA05} and describe the mapped Bose-Hubbard
Hamiltonian in the mean-field approximation. The measure of spin
squeezing and quantum entanglement that we employ is introduced in
Sec.~\ref{s:squeezing}. The results of spin squeezing for
different interaction regimes are presented in
Sec.~\ref{s:results}. Finally, we conclude in
Sec.~\ref{s:conclusion}
\section{Model System}
\label{s:model}
The system we study consists of neutral bosonic atoms with
hyperfine spin $F=1$ in an optical lattice. The optical lattice
results from the ac Stark shifts of standing wave laser fields,
which are dipole coupled to atomic electronic transitions. The
off-resonant coupling induces virtual transitions to electronic
excited states, which upon adiabatic elimination give rise to
level shifts (ac Stark shifts) in the ground state manifold. These
shifts are proportional to the intensity distribution of the laser
light. Additionally two-photon Raman like transitions can couple
any two Zeeman states within the spin-1 ground state manifold,
subject to appropriate polarization selections. In a lattice of ac
Stark shifts from standing waves, the periodic level shift gives
rise to band structures. When the lasers are linearly polarized,
the Zeeman ground state manifold of $(M_F=-1,0,+1)$ remains
degenerate in the lattice. For more general cases of coupling
referred to as the $\Lambda$ or $V$ scheme with suitable
polarizations, two alternate ground states become coupled and will
be denoted as the electronic modes with $\sigma=0$ and
$\sigma=\Lambda$ \cite{KG-PRA04}.

We assume that atoms will remain in the lowest Bloch bands as a
result of the relatively large band gap in comparison to their
kinetic energies. Within this approximation, the atomic field
operator can be expanded in terms of the site localized Wannier
basis. As carefully presented in Ref. \cite{KG-PRA04}, we arrive
at the model Hamiltonian defined on a 1D optical lattice as given
below,
\begin{eqnarray}
\hat{H}_{BH}&=&-\!\!\sum_{\sigma=0,\Lambda}|J_{\sigma}|\sum_{\langle
i,j\rangle} \hat{a}_{\sigma i}^{\dagger} \hat{a}_{\sigma j} \nonumber\\
&&+\sum_{\sigma=0,\Lambda}\frac{U_{\sigma}}{2}\sum_{i}\hat{n}_{\sigma
i}(\hat{n}_{\sigma i}-1) + K\!\sum_{i}\hat{n}_{0i}\hat{n}_{\Lambda i}\! \nonumber \\
&&\!\!-\!
\frac{|P|}{2}\!\sum_{i}(\hat{a}_{0i}^{\dagger}\hat{a}_{0i}^{\dagger}\hat{a}_{\Lambda
i}\hat{a}_{\Lambda i}\!+\!\hat{a}_{\Lambda
i}^{\dagger}\hat{a}_{\Lambda i}^{\dagger}\hat{a}_{0i}\hat{a}_{0i}) \nonumber\\
&&- \delta\sum_{i}\hat{n}_{0i} -
\mu\sum_{\sigma=0,\Lambda}\sum_{i}\hat{n}_{\sigma i},
\label{BH-Ham}
\end{eqnarray}
where $J_{\sigma}$ is the tunnelling parameter, $U_{\sigma}$, $K$,
and $P$ are parameters from the repulsive density-density
interaction of condensed atoms and the spin-exchange interaction.
$\delta$ parameterizes the energy difference between the
electronic internal states $\sigma=0$ and $\sigma=\Lambda$. $\mu$
is the chemical potential, $\hat{a}_{\sigma i}^{\dagger}$ and
$\hat{a}_{\sigma i}$ are respectively creation and annihilation
operators of an atom in mode $\sigma$ at lattice site $i$ and
$\hat{n}_{\sigma i}=\hat{a}_{\sigma i}^{\dagger}\hat{a}_{\sigma
i}$.

As discussed in Ref. \cite{KG-PRA04}, the various parameters of
the above Hamiltonian (\ref{BH-Ham}) can be given in terms of
Wannier spinors, and thus they depend on $\theta$, the angle
between the polarization vectors of the two counter-propagating
linearly polarized laser beams in the lin-$\theta$-lin
configuration of an optical lattice.

In the mean-field approximation \cite{mf} with
$\psi_{\sigma}=\langle \hat{a}_{\sigma j}\rangle$ assumed real
\cite{KG-PRA04}, we substitute
\begin{equation}
\hat{a}_{\sigma i}^{\dagger}\hat{a}_{\sigma j}\approx
\psi_{\sigma}(\hat{a}_{\sigma j}+\hat{a}_{\sigma
i}^{\dagger})-\psi_{\sigma}^2,
\end{equation}
into the Hamiltonian (\ref{BH-Ham}), and arrive at
\begin{eqnarray}
\hat{H}_{BH}^{\rm
MF}&\!\!\!=&\!\!\!-2\!\!\sum_{\sigma=0,\Lambda}\!\!J_{\sigma}
[(\hat{a}_{\sigma}+\hat{a}_{\sigma}^{\dagger})\psi_{\sigma}-\psi_{\sigma}^2]
+\!\!\!
\sum_{\sigma=0,\Lambda}\!\!\frac{U_{\sigma}}{2}\hat{n}_{\sigma}(\hat{n}_{\sigma}\!-1)
\nonumber \\
&&\!\!+ K\hat{n}_{0}\hat{n}_{\Lambda} -
\frac{|P|}{2}(\hat{a}_{0}^{\dagger}\hat{a}_{0}^{\dagger}\hat{a}_{\Lambda}
\hat{a}_{\Lambda}+\hat{a}_{\Lambda}^{\dagger}\hat{a}_{\Lambda}^{\dagger}
\hat{a}_{0}\hat{a}_{0}) \nonumber \\
&&\!\!- \delta\hat{n}_0 - \mu\sum_{\sigma=0,\Lambda}
\hat{n}_{\sigma}, \label{MF-Ham}
\end{eqnarray}
a system of many independent sites. In the above, we have omitted
the site index $i$ so that effectively, the optical lattice model
is reduced to a collection of single site problems.

The basic idea of the mean-field theory (MFT) is to replace the
fluctuating exchange field by an effective average field in an
interacting many-body system. MFT has been found not quite
reliable to describe critical phenomena especially at low
dimensions~\cite{Kadanoff}. In the MFT, one ignores the long range
fluctuations of the order parameter which causes serious errors at
the critical points where the fluctuations dominate the mean
value~\cite{Ginzburg}. Despite these facts, optical lattices have
been extensively studied under a MFT
approach~\cite{Zoller-PRL98,MFT-opticlat}. The interaction term in
the Bose-Hubbard model for the optical lattices, e.g. the
interaction terms in (\ref{BH-Ham}), is due to atom-atom
collisions which can happen only locally, so that it is an on-site
interaction. The sole non-local interaction is the hopping term,
due to tunneling of the atom between the sites. As in the case of
our spin-1 model, MFT treats the spin-spin interactions exactly
while the kinetic coupling is treated approximately.

MFT, as it is used here, based upon the Bogoluibov symmetry
breaking background field theory. Bogoluibov theory is extended to
describe Mott transition by a specific decorrelation approximation
in a consistent MFT~\cite{Mott-MFT}. It can be systematically
improved by considering bigger clusters (2 sites or more) to
employ MFT~\cite{Ferreira-PRB02}. Away from phase boundaries such
an improvement is not essential for us. The fluctuations are due
to collective excitations of the system. Focusing at zero
temperature, and staying away from the phase boundaries, one can
expect the predicted MFT ground states are well established, since
the collective excitations and associated fluctuations would be
weaker in comparison to the mean-field order parameter. In our
investigations we assume the reported ground
states~\cite{KG-PRA04} describe the system in deep quantum phases
away from the phase boundaries.

Similar approach, as is done here, to determine the ground states
have been employed in a more general system that includes external
magnetic field as well~\cite{Chui-PRA03}. MFT cannot be used to
examine spin-spin correlations among different sites for which
effective models can be used~\cite{intersite}. On site spin
fluctuations however can be examined in MFT to reveal any particle
entanglement associated with the reported ground
states~\cite{KG-PRA04}. The question we address here is how the
type and amount of the entanglement among the particles in a
single lattice site would change when the whole lattice system
undergoes quantum phase transitions and the use of MFT is
sufficient for this question.

Beyond zero-temperature, a generalization of the method is given
in Ref.~\cite{Pandit-PRB08}. At non-zero temperatures it is more
crucial to test predictions of MFT for low dimensional systems
against numerical tests. For spin-1 systems, detailed numerical
studies became only very recently
available~\cite{Scalettar-PRL09}; but they have ensured that
similar level of agreement between the MFT predictions and
numerical studies as in spinless systems do occur for the case of
spin-1 systems.

In order to test the validity of MFT that we use in our model, we
studied a simple lattice model having two sites. We used the
Bose-Hubbard Hamiltonian in (\ref{BH-Ham}) and $i$ runs from 1 to
2 with the periodic boundary conditions. The purpose of this
calculation is to investigate the effect of inter-site interaction
on the single-site state. The exact ground state calculations were
done by using those parameter values corresponding to $n=1$ and
$n=2$ Mott phases in the phase diagrams both for the ferromagnetic
and antiferromagnetic regimes in the case of $\theta=0$ and for a
small $\theta$ value. Once the exact two-site ground state is
determined, we calculate the one-site density matrix by tracing
out the other site. Following this procedure, the overlap of
ground states from MFT and exact two-site model can be computed.
Our results show that most of these overlap values are above 0.95,
confirming the success of MFT in calculating one-site ground
states and so that using it to quantify correlations among
particles in a single site.

In general, many-body wave functions are too complicated to
express explicitly, but MFT allows for writing down analytical
wave functions of the ground states and hence one can gain
valuable insights into the quantum correlations in such complex
many-body systems such as spinor condensates in optical lattices.
This insight should serve as a guide even for comprehending
quantum correlations among the lattice sites which require beyond
MFT calculations, but can still be performed through perturbative
examinations of mean field ground states. We hope to investigate
this in near future.

A general spin-1 system is described by the symmetry group SU(3).
In the model considered here, a reduced two-mode description for
the two coupled ground states is represented by a pseudo-spin-1/2
algebra, effectively the isospin subgroup of SU(3)
\cite{KG-PRA04}. The corresponding generators of the SU$(2)$
isospin algebra are given by \cite{KG-PRA04}
\begin{eqnarray}
\hat{T}_1&=&\frac{1}{2}(\hat{a}_{\Lambda}^{\dagger}\hat{a}_{0}+
\hat{a}_{0}^{\dagger}\hat{a}_{\Lambda}), \nonumber \\
\hat{T}_2&=&\frac{\mathrm{i}}{2}(\hat{a}_{\Lambda}^{\dagger}\hat{a}_{0}-
\hat{a}_{0}^{\dagger}\hat{a}_{\Lambda}), \nonumber \\
\hat{T}_3&=&\frac{1}{2}(\hat{a}_{0}^{\dagger}\hat{a}_{0} -
\hat{a}_{\Lambda}^{\dagger}\hat{a}_{\Lambda}), \label{operators}
\end{eqnarray}
in terms of which the mean-field Hamiltonian (\ref{MF-Ham}) can be
expressed as
\begin{eqnarray}
\hat{H}_{BH}^{\rm
MF}&\!\!=&\!\!-2\!\!\sum_{\sigma=0,\Lambda}\!\!J_{\sigma}
[(\hat{a}_{\sigma}+\hat{a}_{\sigma}^{\dagger})\psi_{\sigma}\!-\!\psi_{\sigma}^2]
\nonumber \\
&&\!\!+\frac{U_{\Sigma}}{2}\hat{T}_3^2+(K-|P|)\hat{T}_1^2+(K+|P|)\hat{T}_2^2 \nonumber \\
&&\!\!+\frac{U_{\Sigma}}{8}\hat{n}^2\!\!-\left(\frac{K}{2}
+\mu+\frac{U_{\Sigma}}{4}+\frac{\delta}{2}\right)\hat{n}\nonumber\\
&&\!\!-\left(\frac{\Delta
U}{2}+\delta\right)\hat{T}_3+\frac{\Delta U}{2}\hat{n}\hat{T}_3,
\label{MF-Ham-operators}
\end{eqnarray}
where $\Delta U=U_0-U_{\Lambda}$, $U_{\Sigma}=U_0+U_{\Lambda}$,
and $\hat{n}=\hat{n}_{0}+\hat{n}_{\Lambda}$. Spin dependent
interaction terms in this Hamiltonian emulates that of the
generalized Lipkin-Meshkov-Glick (LMG) model
\cite{lipkin,unanyan}, or its special case of the two-axis
twisting model \cite{Kitagawa-PRA93}. Such models are capable of
generating spin squeezing \cite{Kitagawa-PRA93} and multiparticle
entanglement \cite{Helmerson-PRL01,unanyan}. Our model above,
includes tunnelling and collision effects in addition to the
generalized LMG interaction terms.

When the lattice parameter $\theta=0$, the two modes have the same
energy and $J_0=J_{\Lambda}=J$, $U_0=U_{\Lambda}=U$, $K=U+P$, and
$\delta=0$ \cite{KG-PRA04}. The simplified Hamiltonian
(\ref{MF-Ham}) takes the following form
\begin{eqnarray}
\hat{\cal H}_{\rm af(f)}&=& -2J\sum_{\sigma=0,\Lambda}
[(\hat{a}_{\sigma}^{\dagger}+\hat{a}_{\sigma})\psi_{\sigma}-\psi_{\sigma}^2]
\nonumber\\
&&+2\left(U\hat{T}^2+P\hat{T}_{2(1)}^2\right)+ \alpha\hat{n},
\label{Ham-simp}
\end{eqnarray}
for both antiferromagnetic ($P>0$) and ferromagnetic ($P<0$)
interactions \cite{KG-PRA05}, where we have used
$\hat{T}^2=\hat{n}^2/4+\hat{n}/2$ for the collision interaction in
terms of the total isospin operator $\hat{T}^2$ with
$\alpha=-3U/2-P/2-\mu$. The spin interaction now reduces to that
of a single-axis twisting type \cite{Kitagawa-PRA93}.

The above considerations show that our model allows for the
investigation of effects due to tunnelling and collision on spin
squeezing induced by either the two-axis twisting interaction as
in the generalized LMG model or the single-axis twisting
interaction in the simplified case. In the general case of the LMG
model, particle entanglement thus exists for atoms in the
non-degenerate ground state modes, which become degenerate for the
special case of a lattice with $\theta=0$.
\section{Spin Squeezing and Quantum Entanglement}
\label{s:squeezing}
Squeezed spin states defined by Kitagawa and Ueda
\cite{Kitagawa-PRA93} is widely used in atomic physics, especially
in the context of particle correlation and entanglement. A
criterion was found recently connecting many atom entanglement and
correlation originally from atoms in a Bose-Einstein condensate
(BEC) \cite{Zoller-Nature01}. If the squeezing parameter
\begin{equation}
\xi_{\alpha}^2 =\frac{N(\Delta J_{\alpha})^2}{\langle
J_{\beta}\rangle^2 + \langle J_{\gamma}\rangle^2}, \label{squ-cri}
\end{equation}
is smaller than 1, the two mode bosonic many atom state under
consideration is spin squeezed along the direction of $\alpha$.
$\vec{J}$ is the total pseudo spin operator, while $\alpha$,
$\beta$, and $\gamma$ denote three orthogonal axes. The condition
for $\xi_{\alpha}^2<1$ coincides with the non-separability
criterion of a density matrix for $N$ two state boson
\cite{Zoller-Nature01}. Thus $\xi_{\alpha}^2$ can be used to
measure quantum entanglement in the two state atomic system
discussed above. In our study outlined below, we examine spin
squeezing for the on-site isospin algebra by calculating the
variance and expectation values of the corresponding generators
$T_i$ defined in (\ref{operators}). Our results show clearly the
existence of quantum correlations between atoms on the same
lattice site.

To identify pairwise entanglement in our many-body system, we can
make use of a direct relationship between concurrence
\cite{Wootters}, which is well-known and represents a widely
accepted measure of bipartite entanglement, and spin squeezing
criterion \cite{Sanders}. Thus, we take (\ref{squ-cri}) as an
indicator for two-particle entanglement. We will in addition also
calculate the concurrence and compare the results with the
squeezing parameter (\ref{squ-cri}).

In view of the significant difficulties of measuring spin
squeezing along any arbitrary direction $\alpha$, our
investigation will focus on the simplest case of a single
orthogonal configuration with three fixed axes. Other orthogonal
axes configurations may be sequentially searched for if the
optimal squeezing is to be found. For this aim we only need to
rotate the coordinate system about each of the axes by an angle
$\phi$. For example if the rotation is about the axis-3, $\xi_3^2$
remains the same, while the squeezing parameters for the new
axis-1 and axis-2 become
\begin{eqnarray}
\xi_{1'}^2&\!\!\!=&\!\!N \frac{\Delta T_1^2 \cos^2 \phi + \Delta
T_2^2 \sin^2 \phi - \sin\phi\cos\phi\langle
T_1,T_2\rangle}{\langle T_3\rangle^2 + (\langle T_1\rangle\sin\phi
+ \langle
T_2\rangle\cos\phi)^2} ,\nonumber \\
\xi_{2'}^2&\!\!\!=&\!\!N \frac{\Delta T_1^2 \sin^2 \phi + \Delta
T_2^2 \cos^2 \phi + \sin\phi\cos\phi\langle
T_1,T_2\rangle}{\langle T_3\rangle^2 + (\langle T_1\rangle\cos\phi
- \langle T_2\rangle\sin\phi)^2},\hskip 24pt  \label{squ-rot}
\end{eqnarray}
where $\langle T_i,T_j\rangle = \langle T_i T_j + T_j T_i\rangle -
2\langle T_i\rangle\langle T_j\rangle$.
\section{Results}
\label{s:results}
\subsection{Numerical method}
\label{s:numerics}

The mean-field Bose-Hubbard Hamiltonian in (\ref{MF-Ham}) has been
used to examine the phase transition between the superfluid and
Mott-insulator phases \cite{KG-PRA04}, with $\psi_{\sigma}$
denoting the order parameter for the $\sigma$ mode. The superfluid
phase for the $\sigma$ component is identified with
$\psi_{\sigma}\neq0$. In the superfluid state the tunnelling term
$J_{\sigma}$ is large and dominates the Hamiltonian. As a result
the ground state corresponds to the single particle wave function
of all $\sigma$-type atoms extended over the whole lattice, with
each site being a coherent superposition of Fock number states
\cite{Bloch-Nature02}. In the Mott phase, on the other hand, the
interaction term dominates so that the ground state exhibits
minimal number fluctuation and corresponds to a product of atom
Fock number states at each lattice site, which in turn gives
$\psi_{\sigma}=0$ \cite{Bloch-Nature02}.

We have performed numerical diagonalization of the mean-field
Hamiltonian (\ref{MF-Ham}) by using a set of states expanded in
terms of the product of individual atom number states
\begin{equation}
|\Omega\rangle=\sum_{n_0=0}^N \sum_{n_{\Lambda}=0}^N c_{n_0
n_{\Lambda}} |n_0\rangle |n_{\Lambda}\rangle . \label{basis}
\end{equation}
While performing this diagonalization, two different regimes with
respect to the same parameter $P$ must be carried out. One is for
a positive antisymmetric coupling, with a corresponding
antiferromagnetic ground state, where individual spins are
anti-aligned due to spin-exchange interaction. The other case is
ferromagnetic for a negative spin exchange interaction. In
addition, we explore the dependence of our results on the small,
but non-vanishing lattice parameter $\theta$, which introduces a
spin dependent lattice potential.

We study the parameter regions corresponding to those considered
in Ref. \cite{KG-PRA04}. The values of the parameters in
Hamiltonian (\ref{MF-Ham}), which are needed for numerical
computation, are thus read from the Fig. 1 of Ref.
\cite{KG-PRA04}, with $J/U$ picked to ensure the system have full
access to the $n=2$ Mott regime, but barely enters the $n=3$ Mott
phase. $\theta$ is taken to be small and $\delta$ values used are
for the range of $0\le \theta\le 1$. We study the degenerate
($\theta=0$) and non-degenerate cases ($\theta\ne 0$) separately.
From the initial values of the order parameters $\psi_0$ and
$\psi_{\Lambda}$ we compute the diagonal basis and the
corresponding ground state. This ground state then allows us to
calculate the new order parameters and to compare with the initial
values. This procedure is iterated to reach a self-consistent
solution, with which it becomes straightforward to calculate the
expectation values and the second moments of the operators in
(\ref{operators}).

To conveniently calculate the squeezing parameter $\xi^2$
(\ref{squ-cri}), we use the average total occupation number
$\langle \hat{n}\rangle$ for each type of interactions to label
the different phases instead of relying on the total number of
atoms $N$ (per site). This implicitly assumes that the squeezing
parameter (\ref{squ-cri}) remains a valid criterion of quantum
entanglement even for non-integer occupation numbers such as in
the superfluid phase. This assumption does not introduce any
inconvenience in a Mott phase since the ground state consists of
Fock states with equal total number of particles, {\it i.e.},
definite spin and thus $\langle \hat{n}\rangle$ becomes an
integer. In the superfluid phase, we justify the use of a
non-integer $\langle \hat{n}\rangle$ in the following manner. In
this section, we calculate the squeezing parameter in two
different ways for each case. The first method uses $\langle
\hat{n}\rangle$ directly for the entanglement measure. The second
method is analogous in form, but only uses integer values of
$\langle \hat{n}\rangle$. For the superfluid phase, instead of
talking about separability for states with different total number
of particles, we focus on the subspace $n_0 + n_{\Lambda} = n$
block and investigate its correlation. This becomes a meaningful
measure when the block we use is the one with the nearest integer
total number of particles to $\langle \hat{n}\rangle$. This method
has a similar nature as the superselection rules mentioned in Ref.
\cite{Amico-review} and in Ref. \cite{Vaccaro} since the
projection of the Hilbert space onto a subspace of fixed particle
number is considered. Both
 methods are found to give similar behaviors for the
superfluid and the Mott insulator phases. We provide results from
the first method in our discussion because they respect the
collective nature of the superfluid state and emphasize particle
number fluctuations.

There also exist states for which spin squeezing parameter cannot
be readily used to characterize their correlation properties. An
example is the maximally entangled states (MES) in Ref.
\cite{Micheli-PRA03}, which are not squeezed spin states according
to the criterion in (\ref{squ-cri}). In this case, it is
inadequate to talk about squeezing, since the uncertainty in the
perpendicular components to the mean isospin vector are
meaningless as the denominator for the squeezing measure
(\ref{squ-cri}) vanishes for all axes. In addition, there exist
other states, although whose averaged mean isospin are nonzero,
the expectation values for the two components in the denominator
might vanish, also making the spin squeezing parameter $\xi_i^2$
not well defined. In our studies, we find that these states happen
only in certain Mott phases, where exact wave functions are
available either analytically in the spin \cite{KG-PRA05} or Fock
basis \cite{KG-PRA04}. As such, their quantum entanglement
properties can be discussed directly using other criteria.

In order to quantify the pairwise quantum correlations both in the
superfluid and Mott-insulator regimes, in addition to the
squeezing parameter, we use the well-known criterion called
concurrence \cite{Wootters}. For a given two-party state $\rho$,
this measure is equal to
\begin{equation}
C(\rho)=\max\{0, \lambda_1 -\lambda_2 -\lambda_3 -\lambda_4 \},
\label{concurrence}
\end{equation}
where $\lambda_i$'s are the square roots of the eigenvalues of
$\rho\tilde{\rho}$ in decreasing order where
\begin{equation}
\tilde{\rho}=(\sigma_y \otimes\sigma_y)\rho^{*}(\sigma_y
\otimes\sigma_y) \label{tilderho}.
\end{equation}

For the $n=1$ Mott-insulator phase, this measure is trivial since
there is only one particle present. When it comes to the $n=2$
Mott phase, concurrence clearly quantifies pairwise correlations
between the two atoms at the same lattice site. In the superfluid
phase, the ground state is a superposition of Fock states with
different number of atoms or isospin states with different
isospins, we again focus on the subspace with the nearest integer
total number of particles $n_0 + n_{\Lambda} = n$. If the nearest
integer is smaller than two, then concurrence is zero. If it is
equal to two, the concurrence is simply calculated. When it is
equal to three, the three-particle ground state is symmetrized in
the first quantization picture and we use reduced two-body density
matrix to calculate concurrence.

We report below our investigation of quantum entanglement in our
model system for the two regimes: antiferromagnetic and
ferromagnetic interactions.
\subsection{Ferromagnetic regime}

\begin{figure}[htb]
\includegraphics[width=3.25 in]{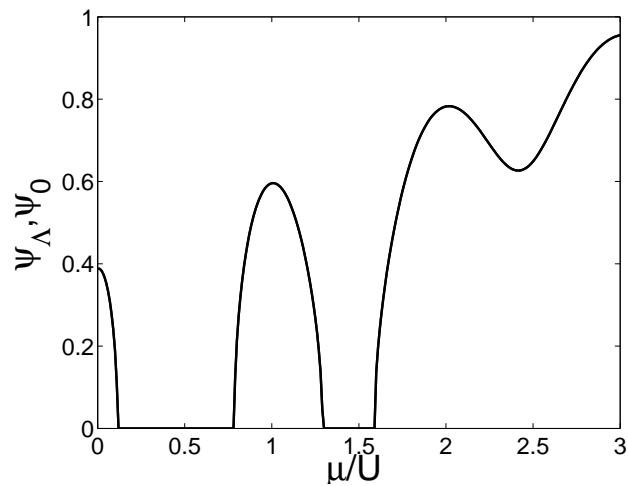}
\caption{The dependence of the order parameters on the value of
$\mu/U$ for $\theta=0$, $J/U=0.455\times 10^{-1}$, and
$P/U=-0.926\times 10^{-2}$ in the ferromagnetic regime. The
vanishing of the order parameters matches closely with the
appearance of Mott-insulator phases for the corresponding
component.} \label{f:ordpar-f}
\end{figure}

For ferromagnetic interaction with $P<0$, for $\theta=0$, and a
fixed $J/U$ value, the dependence of the order parameters $\psi_0$
and $\psi_{\Lambda}$ on the quantity $\mu/U$ is shown in
Fig.~\ref{f:ordpar-f}. We determine the phase of the system for
any $\mu/U$ value by looking at the order parameter of each
component.

Quantum correlations on a single lattice site is evidenced by
evaluating the squeezing parameter (\ref{squ-cri}). In the
superfluid regime, numerical calculations, taking into account the
minimization with respect to coordinate rotations, yield
$\xi_i^2>1$. So there is no particle entanglement in the
superfluid phase of a ferromagnetically interacting system when
$\theta=0$. Although, this situation deserves to be more carefully
analyzed for those values of $\mu/U$ that correspond to
Mott-insulator phases.

The spin squeezing parameter is not defined for the zero particle
($n=0$) ground state $|0,0\rangle$, the trivial case of no
particle entanglement without any particles. When $\theta=0$, the
single particle ($n=1$) ground states $|10\rangle$ and
$|01\rangle$ are degenerate \cite{KG-PRA04} and can be written as
$|g\rangle=\cos{x}|01\rangle+\sin{x}\exp{(\mathrm{i}y)}|10\rangle$,
where $x,y\in [0,2\pi]$ are arbitrary angles, parameterizing the
manifold of the ground state family. We find $\langle
T_1\rangle=(1/2)\sin{2x}\cos{y}$, $\langle
T_2\rangle=(-1/2)\sin{2x}\sin{y}$, and $\langle
T_3\rangle=(-1/2)\cos{2x}$. The spin fluctuations are $\langle
\Delta T_1^2\rangle=(1/4)(1-\sin^2{2x}\cos^2{y})$, $\langle \Delta
T_2^2\rangle=(1/4)(1-\sin^2{2x}\sin^2{y})$, and $\langle \Delta
T_3^2\rangle=(1/4)(\sin^2{2x})$. Thus we obtain $\xi_i^2=1$ in any
direction $i=1,2,3$, for any member of the ground state manifold.
The ground state, expressed in the spin representation
\cite{KG-PRA05}, could be written as an arbitrary superposition of
$|T=1/2,T_1=\pm 1/2\rangle$ spin states. We write
$|g\rangle=|x,y\rangle=\cos{(x/2)}|1/2,1/2\rangle+
\sin{(x/2)}\exp{(iy)}|1/2,-1/2\rangle$ for the ground state in
spin representation. Projection of the total spin onto the $(x,y)$
direction gives the spin component
$S_{x,y}=\sin{x}\cos{y}T_2+\sin{x}\sin{y}T_3+\cos{x}T_1$, whose
eigenstate is $|x,y\rangle$ with eigenvalue $1/2$, such that
$S_{x,y}|x,y\rangle=(1/2)|x,y\rangle$. Such a state is called a
coherent spin state (CSS) \cite{Kitagawa-PRA93}. The ground state
$|g\rangle$ is identified as a pure state of a spin-$1/2$ system,
and as such is a CSS. There exists no other spin to be correlated
with, so that $|g\rangle$ cannot be a squeezed spin state (SSS).
Particles in a CSS are correlated as all spin 1/2 constituents
atoms are pointing along the same direction; although they remain
separable, {\it i.e.}, they are not entangled.

On the other hand, the $n=1$ Mott state could become mode
entangled \cite{duan} for some $\alpha$ and $\beta$. Mode
entanglement is a different concept from particle entanglement
considered here and could be useful for different applications
\cite{duan}. It corresponds to entanglement in the second
quantization picture, while particle entanglement is associated
with the inseparability of the wave function, or density matrix,
in the first quantization.

Similarly, the ground states for the $n=2$ Mott phase are also
degenerate for $\theta=0$. As such they form a manifold
represented by
$|g\rangle=\cos{x}|11\rangle+\sin{x}\exp{(\mathrm{i}y)}|b\rangle$,
where $|b\rangle=(|02\rangle+|20\rangle)/\sqrt{2}$. In this case,
$\langle T_1\rangle=\sin{2x}\cos{y}$ and $\langle
T_{2,3}\rangle=0$. The variances are calculated to be $\langle
\Delta T_1^2\rangle=1-\sin^2{2x}\cos^2{y}$, $\langle \Delta
T_2^2\rangle=\cos^2{x}$, and $\langle \Delta
T_3^2\rangle=\sin^2{x}$. $\xi_1^2$ becomes either undetermined (a
$0/0$ form) or $\infty$ due to vanishing denominators. If we
calculate $\xi_1^2$ after a coordinate rotation by $\phi$ about
the axis-$3$, we find $\xi_{1'}^2$. Minimizing it with respect to
$\phi$, we finally get
$(\xi_{1'}^2)_{\mathrm{min}}=1/(2\sin^2{x}\cos^2{y})$ with its
minimum value at $\phi=\pm \pi/2$. We find
$\xi_3^2=1/(2\cos^2x\cos^2y)$ and $\xi_2^2=1/(2\sin^2x\cos^2y)$.
For some values of $x$ and $y$, $\xi_{2,3}^2<1$ is satisfied.
Hence, particle entanglement exists for some members of the ground
state manifold. This is consistent with the fact that each
degenerate ground state $|11\rangle$ and $|b\rangle$ is particle
entangled. For parameters $x$ and $y$ specifying a dominant
contribution from a particular degenerate component in
$|g\rangle$, particle entanglement is expected. In the spin
representation, the ground state is an arbitrary superposition of
$|T=1,T_1=\pm 1, 0\rangle$. In contrast to the spin-$1/2$ case of
the $n=1$ Mott phase, now SSS (squeezed spin state), where all
particles are entangled, can be found in the ground state family.

When we analyze ferromagnetic regime by calculating the
concurrence in light of the discussion in Sec. \ref{s:numerics},
it is found to be zero for all $\mu/U$ values except those for the
$n=2$ Mott phase. In this case, the ground state is an arbitrary
superposition of two degenerate maximally entangled states, with
the concurrence for each state being equal to one. But the
concurrence for the ground state manifold mentioned above becomes
$C(|g\rangle)=[1-(1/2) \sin^2 (2x)\cos(2y)]^{1/2}$, which is
larger than zero for some values of $x$ and $y$. And this
indicates the possibility of pairwise entanglement for certain
ground states.

\begin{figure}[htb]
\includegraphics[width=3.25in]{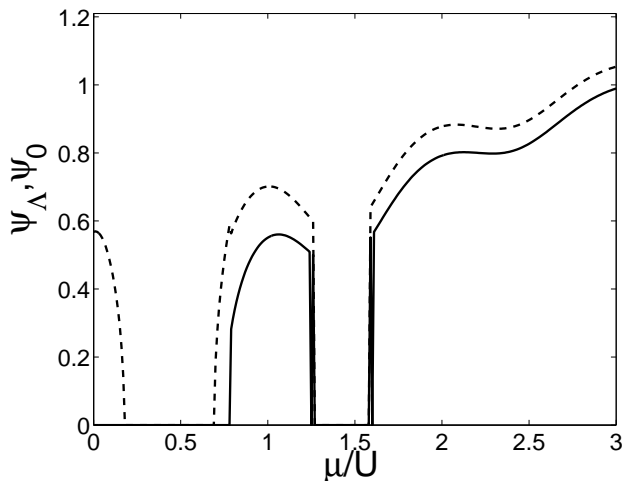}
\caption{The dependence of order parameters for the two modes vs
$\mu/U$ for a small nonzero $\theta$ in the ferromagnetic regime
with $J/U=0.625\times 10^{-1}$, $P/U=-0.926\times 10^{-2}$, and
$\delta/U=0.327\times 10^{-2}$. The solid line denotes
$\psi_{\Lambda}$ while the dashed line refers to $\psi_0$.}
\label{f:ordpar-theta-f}
\end{figure}

Now we look at the situation when $\theta$ takes a small but
nonzero value. In this case the relations $J_0\approx
J_{\Lambda}=J$, $U_0\approx U_{\Lambda}=U$, and $K=U+P$ remain
valid. However, the parameter $\delta$ becomes nonzero, due to the
splitting between the two ground state modes: $\sigma=0$ and
$\sigma=\Lambda$. This causes the dependence of the order
parameters on $\mu/U$ to change as illustrated in
Fig.~\ref{f:ordpar-theta-f}. Note the difference between the order
parameters for the two modes.

For the general case with $\theta\neq 0$, performing minimization
over the axis rotations shows that the optimal squeezing occurs
for the unrotated coordinate axes. By examining the spin squeezing
parameter as a function of $\mu/U$ numerically, we find that
particles are not entangled in the superfluid regime. We thus look
for particle entanglement in the Mott phases.

With a small $\theta$, the degeneracy in the ground states in the
Mott phase is removed. In the $n=1$ Mott-phase, the ground state
becomes $|g\rangle=|n_0=1,n_{\lambda}=0\rangle$. For this state,
the mean spin is along the direction of the axis-$3$ with $\langle
T_3\rangle =1/2$ and $T_{1,2}=0$. The spin fluctuations are given
by $\langle \Delta T_{1,2}^2\rangle=1/4$ and $\langle \Delta
T_3^2\rangle=0$. Employing a rotation by $\phi$ about the
axis-$3$, we find that $\xi_{1',2'}^2=1$, {\it i.e.}, the ground
state is a CSS.

For the $n=2$ Mott-phase, we have a non-degenerate ground state of
the form $|g\rangle=a|02\rangle+b|20\rangle$ \cite{KG-PRA04} with
\begin{eqnarray}
a&=&\left\{\frac{1}{2}\left[1-\frac{\Delta U-2\delta}{\sqrt{\left(
\Delta U-2\delta\right)^2+4P^2}}\right]\right\}^{1/2},\\
b&=&\left\{\frac{1}{2}\left[1+\frac{\Delta U-2\delta}{\sqrt{\left(
\Delta U-2\delta\right)^2+4P^2}}\right]\right\}^{1/2}.
\end{eqnarray}
For such a state, as in the $n=1$ Mott phase, the mean spin is
pointed along the axis-$3$ with $\langle T_{1,2}\rangle=0$ and
$\langle T_3\rangle=b^2-a^2$. Their corresponding fluctuations are
found to be $\langle \Delta T_1^2\rangle=(a+b)^2/2$, $\langle
\Delta T_2^2\rangle=(a-b)^2/2$, and $\langle \Delta
T_3^2\rangle=1-(b^2-a^2)^2$. To determine the optimum noise
reduction and spin squeezing, we minimize over rotations about the
mean spin (axis-$3$) direction by an angle $\phi$. It is
sufficient to consider either one of the rotated $1'$ or $2'$ axes
so that a single rotation angle dependent spin squeezing parameter
$\xi_{\phi}^2$ can be found as
\begin{eqnarray}
\xi_{\phi}^2&=&\frac{1+2ab\cos{2\phi}}{(b^2-a^2)^2}.
\end{eqnarray}
Its minimum occurs at $\phi=\pm\pi/2$ such that
$\xi_{\pm\pi/2}^2=(1-2ab)/(b^2-a^2)^2$. Assuming a small
$\delta/P$, we find $\xi_{\pm\pi/2}^2\sim 1/2+{\mathrm
O}((\delta/P)^2)$, in agreement with numerical calculation
reported in Fig. \ref{f:ordpar-theta-f}. Thus, the ground state is
particle entangled and spin squeezed.

We again calculate the concurrence values for the phases under
consideration. It becomes zero everywhere except $n=2$ Mott phase.
In this situation $C(|g\rangle)=2|ab|$ and for small $\delta/P$
values $C(|g\rangle)\sim 1-{\mathrm O}((\delta/P)^2)$. So that the
results are in complete agreement with those of squeezing
parameter.

\subsection{Antiferromagnetic regime}
\begin{figure}[htb]
\includegraphics[width=3.25in]{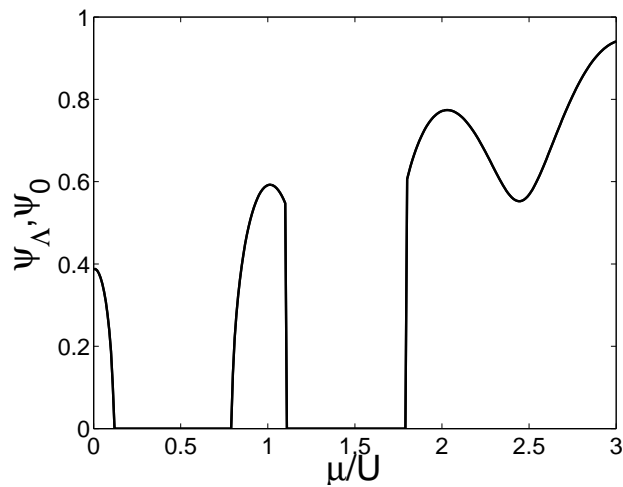}
\caption{The dependence of the order parameters on the values of
$\mu/U$ for $\theta=0$ in the antiferromagnetic regime with
$J/U=0.455\times 10^{-1}$ and $P/U=0.926\times 10^{-2}$. The
nonzero valued order parameters indicate superfluid phases for the
corresponding components.} \label{f:ordpar-af}
\end{figure}

In this case, the atomic interaction parameter $P$ is positive. In
Fig.~\ref{f:ordpar-af}, the order parameters are plotted as a
function of $\mu/U$ at $\theta=0$.

\begin{figure}[htb]
\includegraphics[width=3.25in]{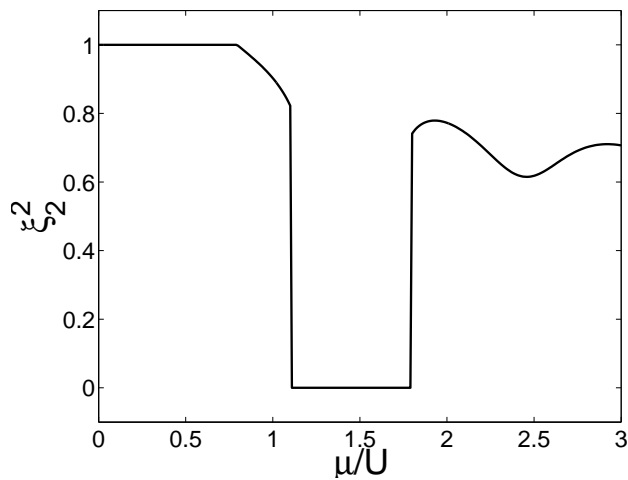}
\caption{The minimum squeezing parameter $\xi_2^2$ for the fixed
axes configuration in the antiferromagnetic regime with
$\theta=0$, $J/U=0.455\times 10^{-1}$, and $P/U=0.926\times
10^{-2}$. $\xi_2^2 <1$ denotes spin squeezing for the axis-$2$.}
\label{f:squ-af}
\end{figure}

Similar to the ferromagnetic regime, we first test the existence
of spin squeezing for $\theta=0$. The corresponding minimum
squeezing parameter, $\xi_2^2$ for the fixed axes is shown in
Fig.~\ref{f:squ-af}. In contrast to the ferromagnetic case,
squeezing is observed for the superfluid phase as well. In
numerical calculations, we also rotate the coordinate system to
see whether correlations can be enhanced for some angles. The
optimum squeezing is found to occur for the fixed axes
configurations.

In the $n=1$ Mott-phase, the ground state is a coherent
superposition of $|10\rangle$ and $|01\rangle$, which identifies a
manifold of any pure state for spin-$1/2$. The only difference
being the quantization axis, it lies along the axis-$2$, instead
of the axis-$1$. Hence our conclusions for the ferromagnetic case
remain applicable. The ground state family is a general CSS and
exhibits no squeezing, although mode entanglement can be present.

The $n=2$ Mott insulator state in the antiferromagnetic case,
however, is significantly different from the ferromagnetic case
considered earlier. It is no longer degenerate as before, and
becomes uniquely determined as
\begin{equation}
|g\rangle = \frac{1}{\sqrt{2}}(|20\rangle + |02\rangle),
\label{state2-af}
\end{equation}
instead. For this special superposition state, the mean isospin
vector becomes zero, with $\langle T_{1,2,3}\rangle=0$. Spin
fluctuations are found to be $\langle \Delta T_{1,3}^2\rangle=1$
and $\langle \Delta T_{2}^2\rangle=0$. Given in the second
quantization form and in the occupation number representation, the
mean number of particles in each mode ($0,\Lambda$) is $1$ and the
state is mode entangled. In the first quantization, denoting
single particle wave functions as $\Psi_{i\sigma}$ for particles
$i=1,2$ in modes $\sigma=0,\Lambda$, $|g\rangle$ is found to
become
$|g\rangle=(1/\sqrt{2})(\Psi_{10}\Psi_{20}+\Psi_{1\Lambda}\Psi_{2\Lambda})$.
This state has maximum quantum correlation among the particles and
can be identified as a MES \cite{Micheli-PRA03}.

\begin{figure}[htb]
\includegraphics[width=3.25in]{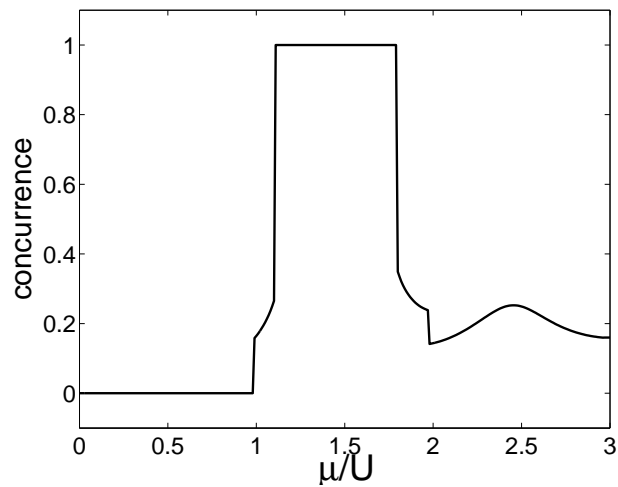}
\caption{Concurrence vs $\mu/U$ in the antiferromagnetic regime
with $\theta=0$. The presence of pairwise entanglement is assured
if the value of concurrence becomes larger than zero.}
\label{f:con-af}
\end{figure}

In order to compare the results measured in terms of the
calculated concurrence, we show in Fig. \ref{f:con-af} the
dependence of concurrence as a function of $\mu/U$.

The presence of particle entanglement in the superfluid phase is
reflected by the nonzero values of concurrence for the
corresponding $\mu/U$ values as shown in Fig. \ref{f:con-af}.
Having a concurrence of one in the $n=2$ Mott phase corresponds to
the presence of a maximally entangled ground state.

\begin{figure}[htb]
\includegraphics[width=3.25 in]{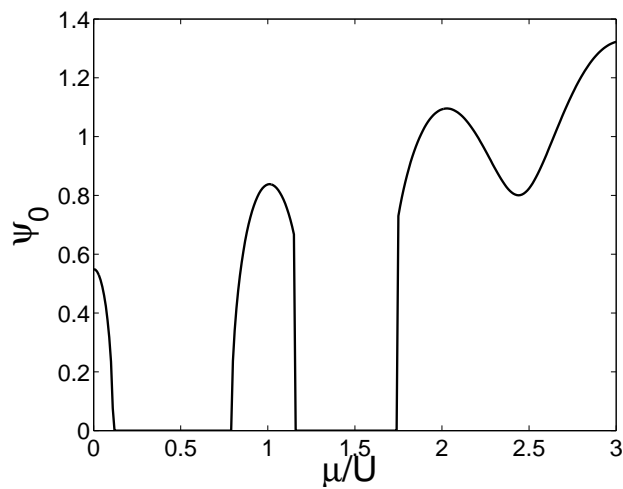}
\caption{The order parameter $\psi_0$ for the antiferromagnetic
regime at a small $\theta$ with $J/U=0.455\times 10^{-1}$,
$P/U=0.926\times 10^{-2}$, and $\delta/U=0.327\times 10^{-2}$.
$\psi_{\Lambda}=0$ for these values of interaction parameters.}
\label{f:ordpar-theta-af}
\end{figure}

As is done previously for the ferromagnetic case, a small nonzero
$\theta$ value can be introduced and the system parameters are
changed accordingly. The corresponding graph for the order
parameters as functions of $\mu/U$ are shown in
Fig.~\ref{f:ordpar-theta-af}.

\begin{figure}[htb]
\includegraphics[width=3.25 in]{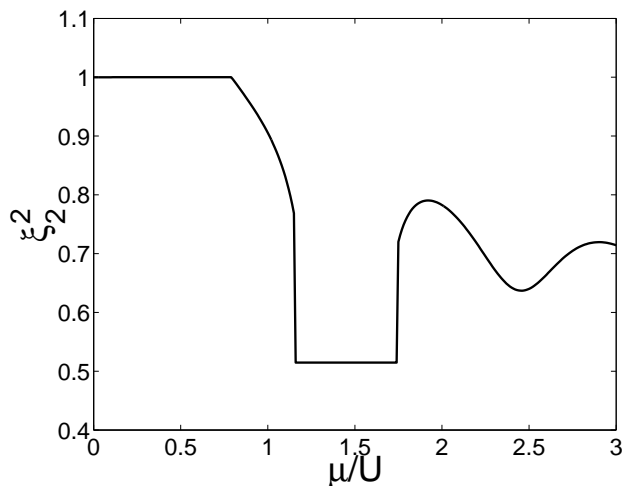}
\caption{The minimum squeezing parameter $\xi_2^2$ in the
antiferromagnetic regime at a small $\theta$ in the fixed
coordinate system with $J/U=0.455\times 10^{-1}$, $P/U=0.926\times
10^{-2}$, and $\delta/U=0.327\times 10^{-2}$.}
\label{f:squ-theta-af}
\end{figure}

Following the earlier procedure, the minimum squeezing parameter
$\xi_2^2$ is also plotted against $\mu/U$, with the optimized
values, corresponding to the fixed coordinate system shown in
Fig.~\ref{f:squ-theta-af}.

As in the case of $\theta=0$, spin squeezing is found to exist for
the superfluid phase almost with the same strength. On the other
hand, although spin squeezing is detected in the $n=2$ Mott phase,
it is reduced with a nonzero $\theta$. The corresponding ground
state for the $n=2$ Mott phase is the same as in the ferromagnetic
case. The MES of the $\theta=0$ case for the antiferromagnetic
interaction becomes a partially entangled state when a small
nonzero $\theta$ is introduced.

%
%
\begin{figure}[htb]
\includegraphics[width=3.25 in]{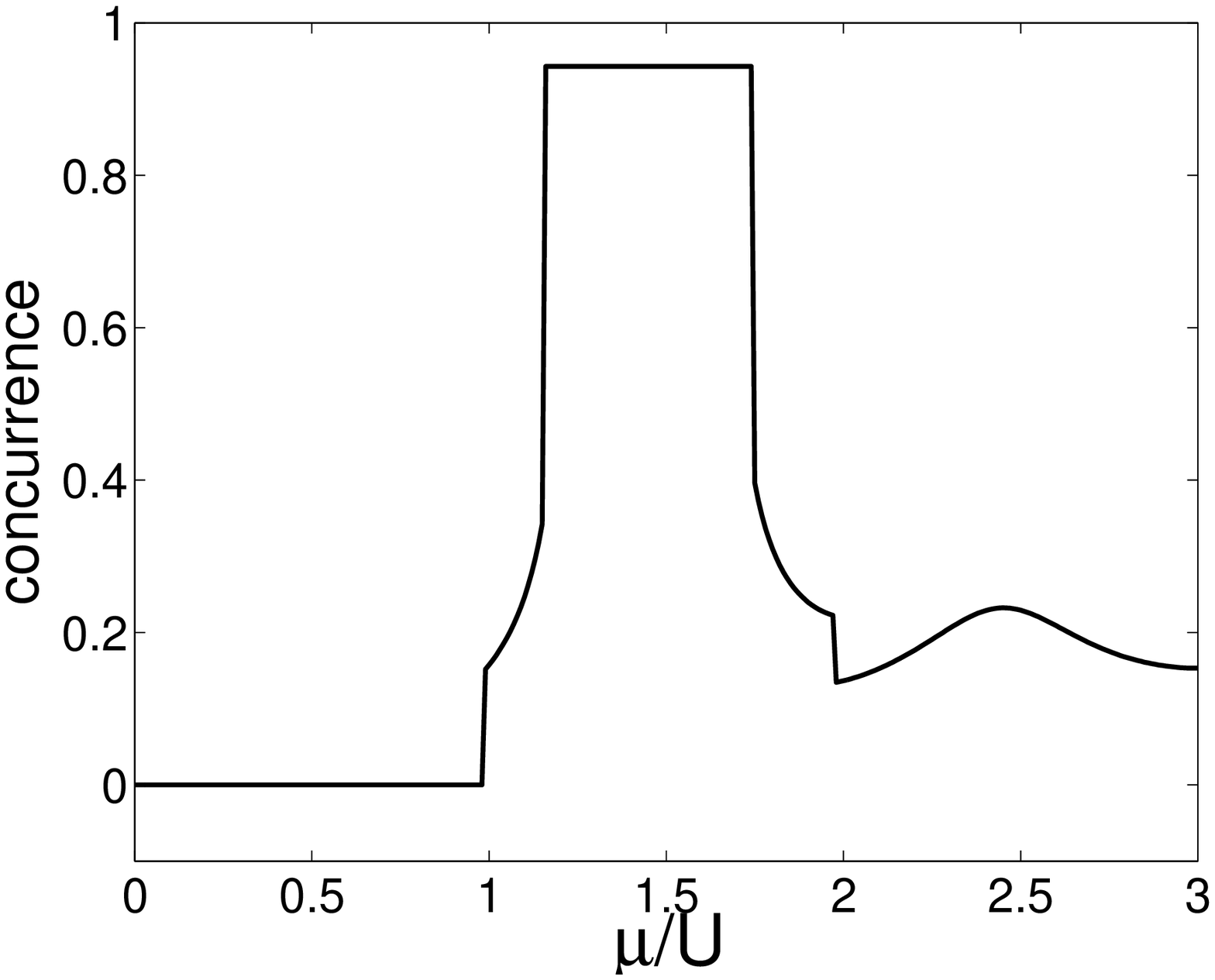}
\caption{Concurrence at a small but nonzero $\theta$ in the
antiferromagnetic regime. It is seen that part of the superfluid
phase contains entangled particles.} \label{f:con-theta-af}
\end{figure}

The results from the calculated concurrence as shown in Fig.
\ref{f:con-theta-af} are in complete agreement with those from the
squeezing parameter. Squeezing is present in the superfluid phase
and the maximal entanglement in the $n=2$ Mott phase becomes
partially entangled with the introduction of a small but nonzero
$\theta$.

\section{Conclusion}
\label{s:conclusion}
In summary, we have investigated quantum correlations between
spin-1 bosons with coupled ground states in optical lattices. Both
ferromagnetic and antiferromagnetic interactions are considered
based on a model, initially developed in Ref. \cite{KG-PRA04},
that we believe can be readily adopted to current experimental
systems. In addition to characterizing quantum correlations in
various quantum phases in terms of coherent and squeezed spin
states, and addressing both particle and mode entanglement, the
role of lattice parameter in the familiar lin-$\theta$-lin
configuration is examined.

We have shown that for ferromagnetic interactions isospin
squeezing (or multi-particle entanglement) is absent in the
lattice model of spin-1 bosons in the superfluid phase. The one
particle Mott phase is in fact in a CSS, which is not particle
entangled, although it displays significant mode entanglement. The
two particle Mott state may contain SSS and entangled particles,
if one of the degenerate component in the ground state manifold is
made to dominant. It can be steered into a particle entangled
state by introducing a nonzero $\theta$ to lift the degeneracy,
while the CSS of the $n=1$ Mott phase or the superfluid phase
remains unentangled. The path to quantum entanglement is through
the well known single axis twisting type nonlinear
interaction~\cite{Kitagawa-PRA93} for the degenerate ($\theta=0$)
case. With a nonzero $\theta$, quantum entanglement is generated
from a generalized LMG interaction, which includes a two-axis
twisting type of spin-spin nonlinear interaction.

For antiferromagnetic interactions, spin squeezing and particle
entanglement is found in both the $n=2$ Mott and superfluid
phases. In the $n=2$ Mott state we find maximally entangled
particles. Introducing a nonzero $\theta$ reduces this to a
partially entangled state, and thus decreases particle
correlations.

We compared the results of the squeezing parameter (\ref{squ-cri})
with those of the concurrence (\ref{concurrence}) for each type of
interaction and lattice configuration. They are in complete
agreement in demonstrating the presence or absence of entanglement
for the different phases.

For the system under consideration, we have investigated the
potential ground states and the corresponding quantum correlations
via examining entanglement/squeezing properties. Depending on the
interaction parameters of the system, abrupt changes may occur if
one considers the behavior of entanglement properties. One can
introduce symmetry breaking perturbations to the Hamiltonian
(\ref{MF-Ham-operators}) to remove the degeneracy present in the
various ground states. This can be done via including magnetic
fields and Raman pulses with which adjustments to the ground state
populations in any particular spin components can be made
\cite{Bergmann-review}. As a specific example, generation of a
coherent superposition of degenerate states (in this case Zeeman
sublevels $M_F =\pm 1$) by stimulated Raman adiabatic passage
scheme is demonstrated experimentally in Ref.
\cite{Bergmann-spin1}.

\ack

M. \"O. O. is supported by T\"UB{\. I}TAK-KAR{\. I}YER Grant No.
104T165 and a T\"UBA-GEB{\. I}P grant. B. \"{O}. would like to
acknowledge the Scientific and Technical Research Council of
Turkey (T\"UB\.ITAK) for financial support. L.Y. is supported by
US NSF.


\end{document}